\documentclass[a4paper]{jpconf}
\usepackage{graphicx}
\begin{document}
\title{New experimental limit on Pauli Exclusion Principle violation by
electrons 
(the VIP experiment)}

\author{C.~Curceanu~(Petrascu)$^{1,2}$,
S.~Bartalucci$^1$, S.~Bertolucci$^1$, M.~Bragadireanu$^{1,2}$, 
M.~Cargnelli$^3$, M.~Catitti$^1$,  
S.~Di Matteo$^1$, J.-P.~Egger$^4$, C.~Guaraldo$^1$, M.~Iliescu$^{1,2}$, 
T.~Ishiwatari$^3$, M.~Laubenstein$^5$, J.~Marton$^3$, E.~Milotti$^6$, 
D. Pietreanu$^{1,7}$, T. Ponta$^2$, D.L. Sirghi$^{1,2}$, F.~Sirghi$^{1,2}$, 
L. Sperandio$^1$, O. Vazquez Doce$^1$, E.~Widmann$^3$, J.~Zmeskal$^3$}

\address{$^1$INFN, Laboratori Nazionali di Frascati, C. P. 13, Via E. Fermi 40, I-00044, Frascati (Roma), Italy}

\address{$^2$'Horia Hulubei' National Institute of Physics and
 Nuclear Engineering, Str. Atomistilor no. 407, P.O. Box MG-6, 
Bucharest - Magurele, Romania}

\address{$^3$Stefan Meyer Institute for Subatomic Physics,
 Boltzmanngasse 3, A-1090 Vienna, Austria}

\address{$^4$Institute de Physique, Universit\'e de Neuch\^atel,1 rue A. -L. Breguet, CH-2000 Neuch\^atel, Switzerland}

\address{$^5$Laboratori Nazionali del Gran Sasso, 
S.S. 17/bis, I-67010 Assergi (AQ), Italy}

\address{$^6$Dipartimento di Fisica, Universit\`{a} di Trieste and INFN-- Sezione di Trieste, Via Valerio, 2, I-34127 Trieste, Italy}

\address{$^7$UMF Carol Davila, 8 Blv. Eroilor Sanitari, Bucharest, Romania}

\ead{petrascu@lnf.infn.it}

\begin{abstract}
The Pauli Exclusion Principle is one of the basic principles 
of modern physics and is at the very basis of our understanding of matter: 
thus it is fundamental importance to test the limits of its validity. Here 
we present the VIP (Violation of the Pauli Exclusion Principle) experiment, 
where we search for anomalous X-rays emitted by copper atoms in a conductor: 
any detection of these anomalous X-rays would mark a Pauli-forbidden
 transition. VIP is currently taking data at the Gran Sasso underground 
 laboratories, and its scientific goal is to improve by at least four 
 orders of magnitude the previous limit on the probability of Pauli violating 
 transitions, bringing it into the $10^{-29\div-30}$ region. First experimental 
 results, together with future plans, are presented.
\end{abstract}

\section{Introduction}

The Pauli exclusion principle (PEP), which plays a fundamental role in
our understanding of many physical and chemical phenomena,
from the periodic table of elements, to the electric conductivity
in metals, to the degeneracy pressure (which makes white
dwarfs and neutron stars stable), is a consequence of the
spin-statistics connection \cite{pauli}. 
Although the principle has been spectacularly confirmed by the number
and accuracy of its predictions, its foundation lies deep in the
structure of quantum field theory and has defied all attempts to
produce a simple proof, as nicely stressed by Feynman \cite{feynman}.
Given its basic standing in quantum theory, it seems appropriate
to carry out precise tests of the PEP validity and, indeed,
mainly in the last 15-20 years, several experiments have been performed
to search for possible small 
violations \cite{bernabei,borexino,hilborn,nemo,nolte,tsipenyuk}. 
Often, these experiments
were born as by-products of experiments with a different
objective (e.g., dark matter searches, proton decay, etc.), and
most of the recent limits on the validity of PEP have been obtained
for nuclei or nucleons.

 In 1988 Ramberg and Snow \cite{ramberg} performed a dedicated experiment, searching for
anomalous X-ray transitions, that would point to a small violation
of PEP in a copper conductor. The result of the experiment
was a probability $\frac{\beta^{2}}{2}\textless  1.7\times10^{-26}$ that the PEP is violated by electrons.
The VIP Collaboration set up an improved version of the Ramberg and
Snow experiment, with a higher sensitivity apparatus \cite{vip}. Our
final aim is to lower the PEP violation limit for electrons by 
$3 \div 4$ orders of magnitude, by using high resolution Charge-Coupled
Devices (CCDs), as soft X-rays detectors 
\cite{culhane,egger,fiorucci,varidel,kraft}, 
and decreasing the effect of background by a careful choice of the
materials and shielding the apparatus from cosmic rays by setting it in 
the LNGS underground laboratory of the Italian Institute for Nuclear Physics
(INFN).

In the next sections we describe the experimental setup, the
outcome of a first measurement performed in the Frascati
National Laboratories (LNF) of INFN in 2005, along with a
brief discussion on the preliminary results
obtained running VIP at the Gran Sasso National Laboratory (LNGS) of INFN.

\section{The VIP experiment}

The idea of the VIP ({\underline VI}olation of the {\underline P}auli 
Exclusion Principle)
experiment was originated by the 
availability of the DEAR (DA$\Phi$NE Exotic Atom Research) 
setup, 
after it had successfully completed its program at the 
DA$\Phi$NE collider  at LNF-INFN \cite{DEAR}. 
DEAR  used Charge-Coupled Devices (CCDs) as detectors in 
order to measure exotic atoms (kaonic nitrogen and 
kaonic hydrogen) X-ray transitions.
CCDs are almost ideal detectors for X-rays measurement, due to their excellent
background rejection capability, based on pattern recognition, and
to their good energy resolution (320 eV FWHM at 8 keV in the present measurement).


\paragraph{Experimental method}

The experimental method, originally described in \cite{ramberg}, consists
in the introduction of new electrons into a copper strip, by
circulating a current, and in the search for X-rays resulting from
the forbidden radiative transition 
that occurs if one
of the new electrons is captured by a copper atom and cascades
down to the 1s state already filled by two electrons with opposite
spins. The energy of this transition would differ from the normal $K_{\alpha}$ transition by
about 300 eV (7.729 keV instead of 8.040 keV) \cite{sperandio}, providing an
unambiguous signal of the PEP violation. The measurement alternates
periods without current in the copper strip, in order to
evaluate the X-ray background in conditions where no PEP violating
transitions are expected to occur, with periods in which current flows in the conductor, thus providing ``fresh'' electrons,
which might possibly violate PEP.

\begin{figure}[h]
\centering
\includegraphics[height=3.8in]{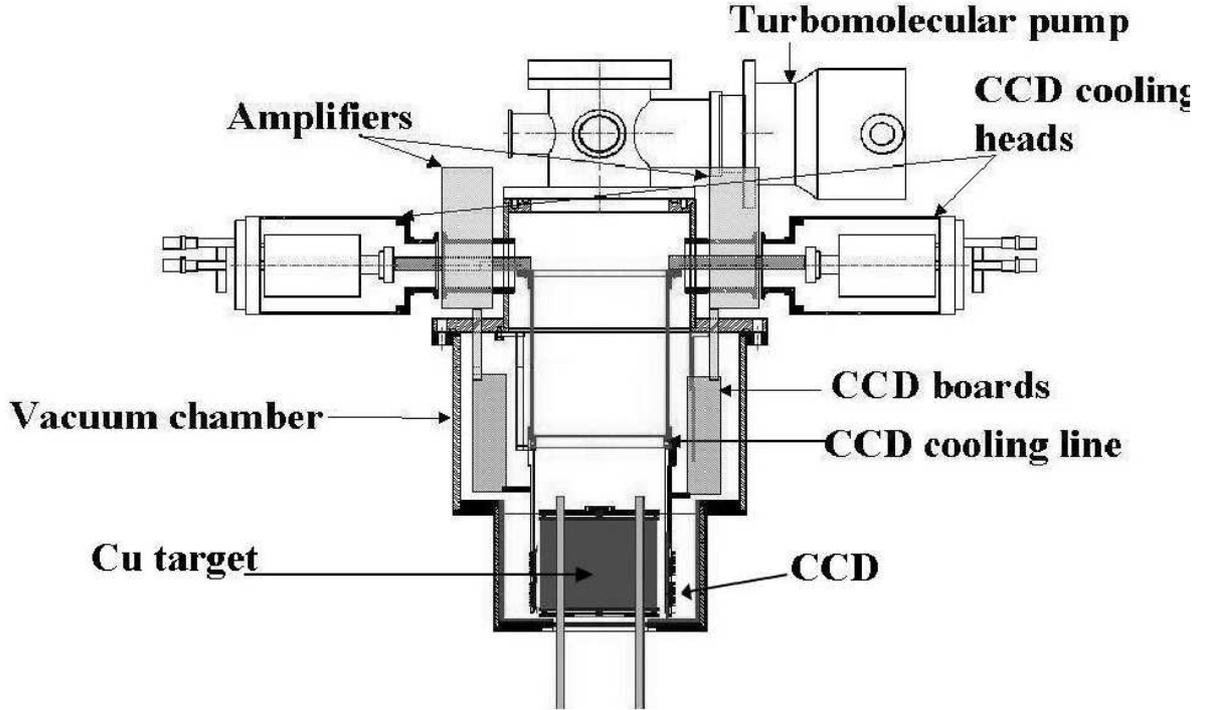}
\caption{The VIP setup - schematic view}\label{setup}
\end{figure}

\paragraph{The VIP setup}
The VIP setup consists of a copper cylinder with 45 mm radius, 50 $\mu$m thickness, 
88 mm height,
surrounded by 16 equally spaced CCDs of type 55 made by EEV \cite{eev}. The CCDs are
at a distance of 23 mm from the copper cylinder, grouped in units of two chips, one
above the other. The setup is enclosed in a vacuum chamber, and the CCDs are
cooled to about 168 K by the use of a cryogenic system. The
current flows in the thin cylinder made of ultrapure copper foil at the bottom of the
vacuum chamber. The CCDs surround the cylinder and are supported by cooling
fingers which are projected from the cooling heads in the upper part of the chamber. The
CCDs readout electronics is just behind the cooling fingers; the signals are sent to
amplifiers on the top of the chamber. The amplified signals are read out by ADC boards in the data acquisition computer.
More details on the CCD-55 performance, as well on the analysis method used
to reject background events, can be found in references \cite{DEAR,ishiwatari}. 
A schematic view of the setup is shown in fig. \ref{setup}.


\section{First VIP experimental results}
\begin{figure}[ht]
\centering
\includegraphics[height=2.8in]{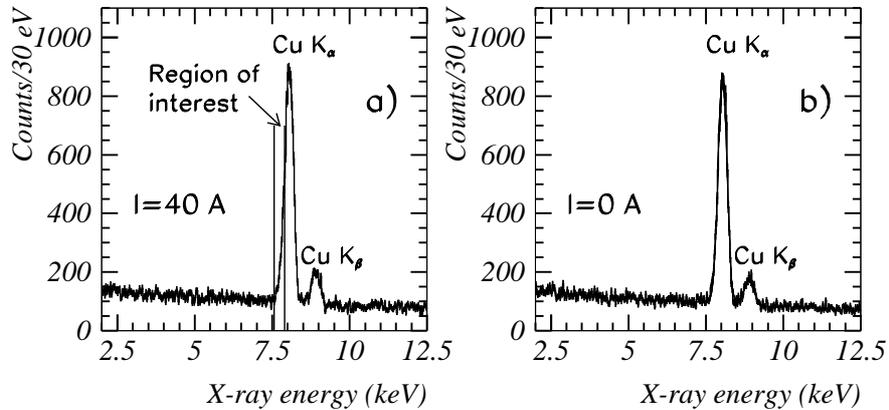}
\caption{Energy spectra with the VIP setup in laboratory: (a) with current (I = 40 A); (b) without current (I = 0).}\label{spectru1}
\end{figure}

\begin{figure}[ht]
\centering
\includegraphics[height=2.7in]{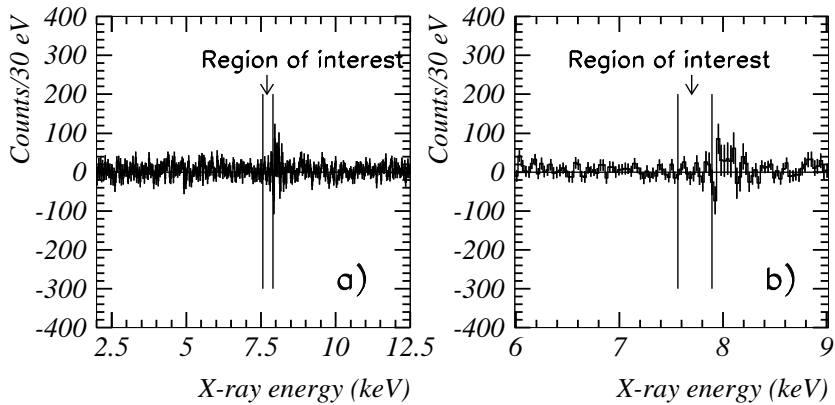}
\caption{Subtracted energy spectra in the Frascati measurement, current minus no-current, giving
the limit on PEP violation for electrons: a) whole energy range; b) expanded view in the region
of interest (7.564 - 7.894 keV). No evidence for a peak in the region of interest is found.}\label{spectru2}
\end{figure}

The VIP setup is presently taking data in the low-background Gran Sasso underground
laboratory of INFN. Before installation in the Gran Sasso laboratory, 
it was first prepared and tested in the LNF-INFN laboratory, where measurements
were performed in the period 21 November - 13 December 2005. Two types
of measurements were performed:
\begin{itemize}
\item 14510 minutes (about 10 days) of measurements with a 40 A current circulating
in the copper target;
\item 14510 minutes of measurements without current.
\end{itemize}
CCDs were read-out every 10 minutes. The resulting energy calibrated
X-ray spectra are shown in figure \ref{spectru1}.
These spectra include data from 14 CCDs out
of 16, because of noise problems in the remaining 2.
Both spectra, apart of
the continuous background component, display clear Cu $K_{\alpha}$ and $K_{\beta}$ lines due to X-ray fluorescence caused by the cosmic ray background and natural radioactivity. No other lines
are present and this reflects the careful choice of the materials used in the setup,
as for example the high purity copper and high purity aluminium, the last one
with $K$-complex transition energies below 2 keV.
The subtracted spectrum is shown in Figure \ref{spectru2} a) (whole energy scale) and b) (a
zoom on the region of interest). Notice that the subtracted spectrum is normalized
to zero within statistical error, and is structureless. This not only yields an upper
bound for a violation of the Pauli Exclusion Principle for electrons, but also confirms
the correctness of the energy calibration procedure.

To determine the experimental limit on the probability that PEP is violated for electrons, $\frac{\beta^{2}}{2}$,
from our data, we used the same arguments of Ramberg and Snow: see references
\cite{ramberg} and  \cite{vipart} for details of the analysis. The obtained value is:
\begin{eqnarray}
\frac{\beta^{2}}{2} \textless 4.5
\times 10^{-28}\quad\quad 
\end{eqnarray}
We have thus improved the limit obtained by Ramberg and Snow by a factor about 40.

\section{Preliminary LNGS updated results}
\begin{figure}[ht]
\centering
\includegraphics[height=3.8in]{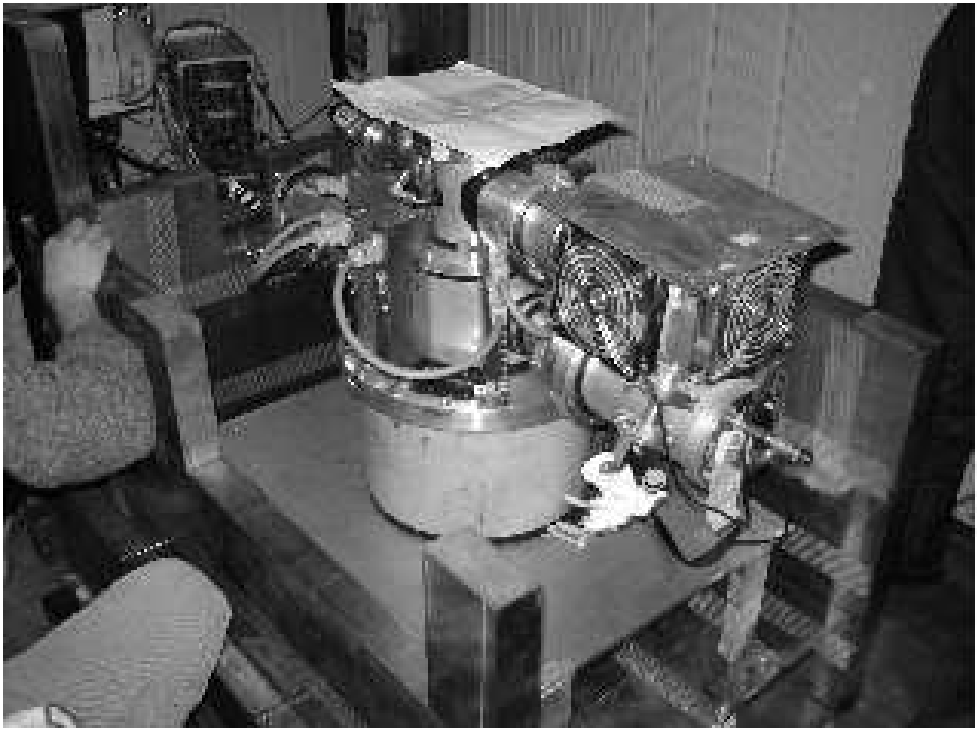}
\caption{Installation of the VIP setup in Gran Sasso underground laboratory}\label{gransassowork}
\end{figure}

\begin{figure}[ht]
\centering
\includegraphics[height=3.8in]{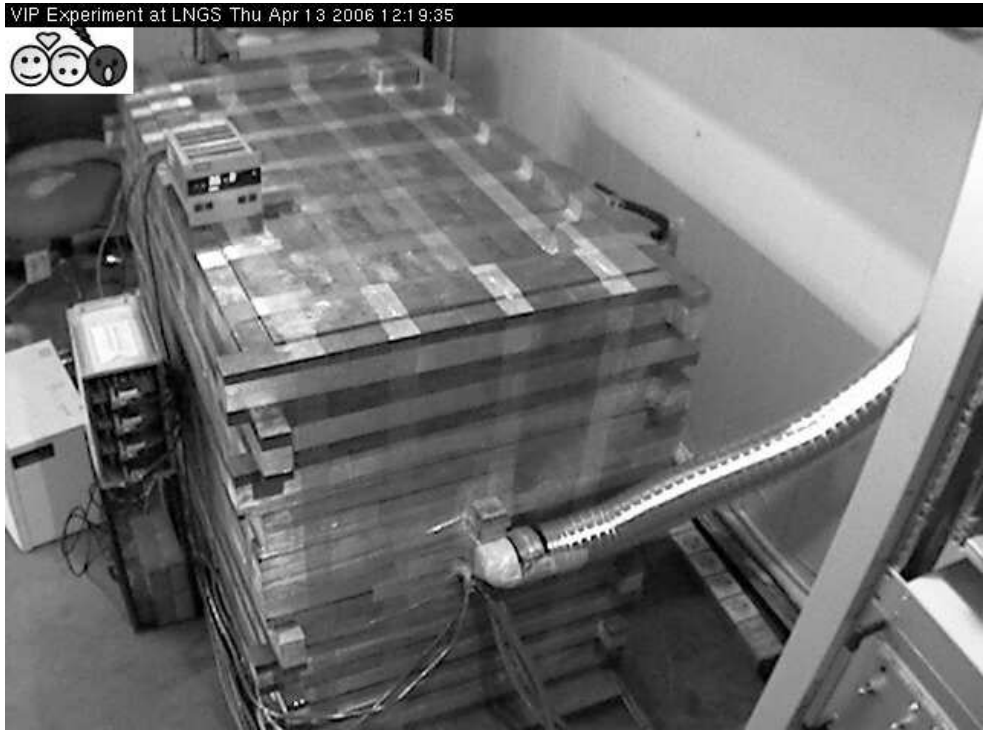}
\caption{The VIP setup taking data at the Gran Sasso underground laboratory}\label{lngs}
\end{figure}

In order to reduce the background, the apparatus is currently installed in the
LNGS underground laboratory, to reduce cosmic rays interactions,
while the effects of natural radioactivity are moderated
by a massive shield built by low activity materials, as it is shown in 
figures \ref{gransassowork} and
\ref{lngs}). 
At the time of writing, about one year of data have been analyzed, 
in the above-mentioned acquisition conditions. Our data consist of two 
periods of data acquisition:     
\begin{itemize}
\item{} about 236005 minutes 
of measurements with a 40 A current 
in the copper target;
\item{} about 172685 minutes of measurements without 
current.
\end{itemize}
\noindent
where CCDs were read-out every 10 minutes. Performing the analysis 
as described in the previous section, from the subtracted spectra we find the new 
preliminary value for the violation parameter:
\begin{eqnarray}
\frac{\beta^{2}}{2} \leq 6.0
\times 10^{-29}\hspace{0.5cm}
\end{eqnarray} 

\begin{flushleft} 
improving the limit obtained by Ramberg and Snow by a factor about 250. 
Today this is the best value ever reached on the probability of PEP violation 
for the electrons. 
\end{flushleft}

\section{Conclusions and perspectives}

A new measurement of the PEP violation
limit for electrons is being performed by the VIP experiment. 
The search of a tiny violation is based
on a measurement of PEP violating X-ray transitions in copper,
under a circulating 40 A current. A new limit for the
PEP violation for electrons was established: $4.5 \times 10^{-28}$, lowering
by about two orders of magnitude the previous one \cite{vipart}.
We have installed VIP at LNGS in spring 2006 and started the data-taking 
in reduced background condition.
After one year of data taking the LNGS preliminary limit for the
PEP violation for electrons was lowered to $6.0 \times 10^{-29}$, 
 about three orders of magnitude better than the Ramberg and Snow
  experiment \cite{ramberg}.
Data acquisition will continue at LNGS with the goal of bringing the limit on 
PEP violation for electrons
into the 10$^{-30}$ region, which is of particular interest \cite{duck}
for all those theories related to possible PEP violation coming
from new physics.

\vspace{4cm}



\end{document}